\documentclass[pdflatex,sn-vancouver,Numbered]{sn-jnl}

\usepackage{graphicx}%
\usepackage{multirow}%
\usepackage{amsmath,amssymb,amsfonts}%
\usepackage{amsthm}%
\usepackage{mathrsfs}%
\usepackage[title]{appendix}%
\usepackage{xcolor}%
\usepackage{textcomp}%
\usepackage{manyfoot}%
\usepackage{booktabs}%
\usepackage{algorithm}%
\usepackage{algorithmicx}%
\usepackage{algpseudocode}%
\usepackage{listings}%

\begin{document}

\title[Promises and challenges of generative artificial intelligence for human learning]{Promises and challenges of generative artificial intelligence for human learning}

\author[1]{\fnm{Lixiang} \sur{Yan}}\email{lixiang.yan@monash.edu}
\author*[2,3]{\fnm{Samuel} \sur{Greiff}}\email{samuel.greiff@gmail.com}
\author[2]{\fnm{Ziwen} \sur{Teuber}}\email{ziwen.teuber@uni.lu}
\author*[1]{\fnm{Dragan} \sur{Gašević}}\email{dragan.gasevic@monash.edu}



\affil[1]{\orgdiv{Faculty of Information Technology}, \orgname{Monash University}, \orgaddress{\street{20 Exhibition Walk}, \city{Clayton}, \postcode{3168}, \state{Victoria}, \country{Australia}}}

\affil[2]{\orgdiv{Department of Behavioral \& Cognitive Science}, \orgname{University of Luxembourg}, \orgaddress{\street{11, porte des sciences}, \city{Esch}, \postcode{4366}, \country{Luxembourg}}}

\affil[3]{\orgdiv{Department of Educational Psychology}, \orgname{Goethe-University Frankfurt}, \orgaddress{\street{Theodor-W.-Adorno-Platz 6}, \city{Frankfurt}, \postcode{60323}, \country{Germany}}}

\abstract{
Generative artificial intelligence (GenAI) holds the potential to transform the delivery, cultivation, and evaluation of human learning. This Perspective examines the integration of GenAI as a tool for human learning, addressing its promises and challenges from a holistic viewpoint that integrates insights from learning sciences, educational technology, and human-computer interaction. GenAI promises to enhance learning experiences by scaling personalised support, diversifying learning materials, enabling timely feedback, and innovating assessment methods. However, it also presents critical issues such as model imperfections, ethical dilemmas, and the disruption of traditional assessments. Cultivating AI literacy and adaptive skills is imperative for facilitating informed engagement with GenAI technologies. Rigorous research across learning contexts is essential to evaluate GenAI's impact on human cognition, metacognition, and creativity. Humanity must learn with and about GenAI, ensuring it becomes a powerful ally in the pursuit of knowledge and innovation, rather than a crutch that undermines our intellectual abilities.
}


\maketitle

\section{Main}\label{sec-main}

Human learning is a journey that shapes minds, fosters innovation, and builds the foundations of society. Beyond merely acquiring knowledge and skills, learning is a path towards fostering critical thinking, creativity, collaboration, and social cohesion. By nurturing the ability to question, analyse, and innovate, learning empowers individuals to navigate complex challenges and contribute to societal progress. While education encompasses formalised systems that structure learning processes, learning represents the dynamic and personal process that occurs within this framework (Table \ref{tab:learning}).

The history of human learning presents a narrative of continuous evolution and adaptation to technological breakthroughs. For example, the printing press democratised access to knowledge and opened the opportunity of learning to many, while the Internet and digital technologies transformed information dissemination and collaborative learning across time and space. In this continuum of innovation, recent advancements in artificial intelligence (AI) present another transformative opportunity to rethink learning processes and educational methodologies \cite{gavsevic2023empowering}. 

Generative AI (GenAI) technologies, such as large language models (LLMs) and diffusion models (Table \ref{tab:ai}), have shown promise in automating various learning tasks \cite{yan2023practical}, delivering feedback on human efficacy \cite{dai2023can}, outperforming average students in reflective writing \cite{li2023can}, innovating conversational assessments \cite{yildirim2023conversation}, creating dynamic learning resources \cite{mazzoli2023enhancing}, and supporting multimedia learning \cite{vartiainen2023using}. However, these technologies also present challenges and ethical considerations that could outweigh their benefits \cite{yan2023practical, kasneci2023chatgpt}. One major concern is the digital divide, where unequal access to these powerful technologies can exacerbate existing inequalities in learning opportunities \cite{pontual2020applications}. Additionally, overreliance on GenAI may negatively impact learners' agency, critical thinking, and creativity, warranting caution \cite{darvishi2023impact}.

\begin{figure}[ht]
    \centering
      \includegraphics[width=.99\textwidth]{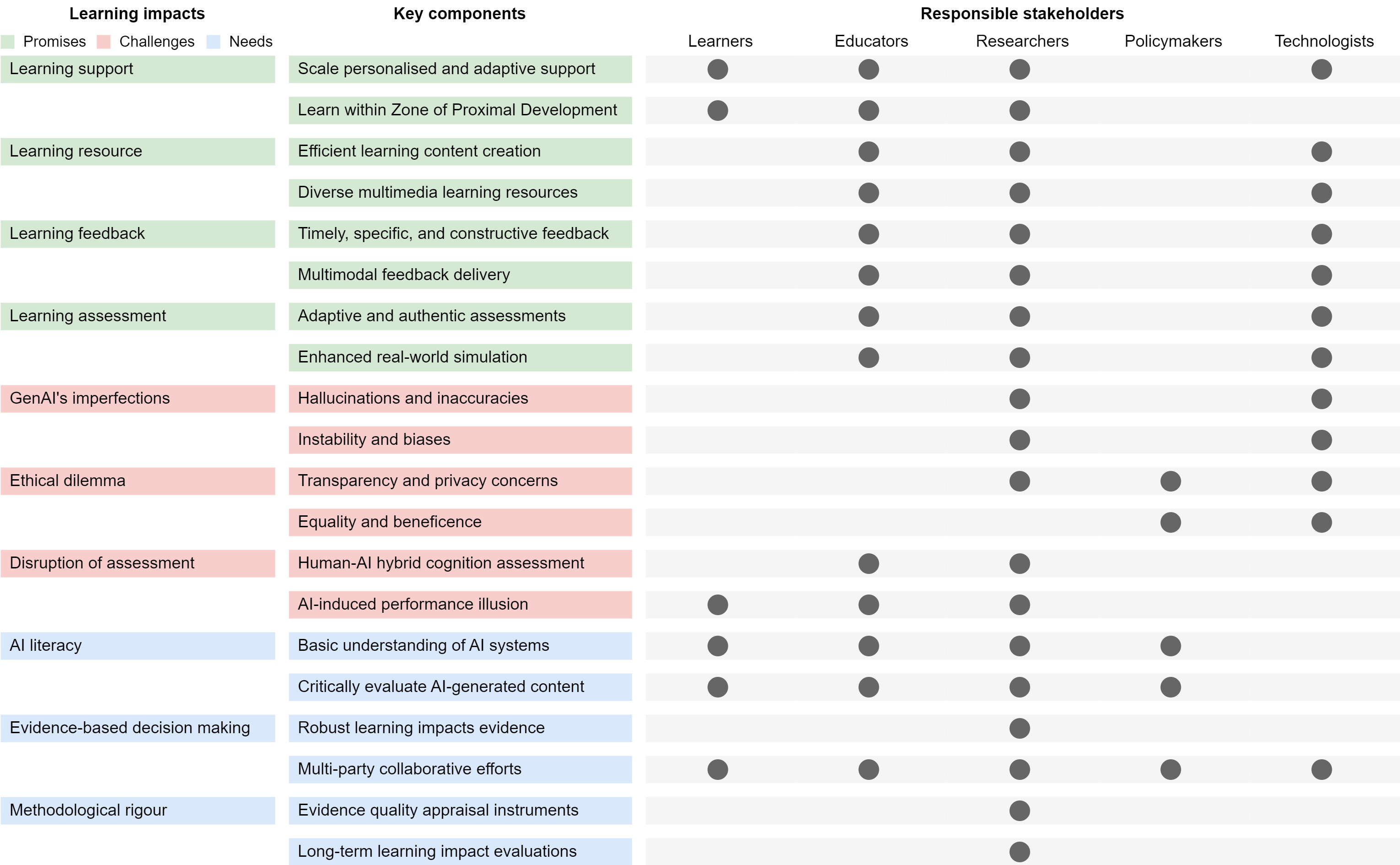}
    \caption{\textbf{Overview of the impacts of generative artificial intelligence on human learning.} The left side of the figure lists various learning impacts, which are categorised into promises (green), challenges (red), and needs (blue). The middle column presents key components associated with each learning impact. These components detail specific aspects that need to be addressed or leveraged to use generative AI as a tool for learning. The matrix on the right shows the five main groups involved in implementing these key components: learners, educators, researchers, policymakers, and technologists. The dots in each column indicate that the relevant group needs to make a substantive contribution to achieving the goals of the key component in the corresponding row.}
    \label{fig: overview}
\end{figure}

Consequently, it is essential to balance technological advancement and human-centred values in learning. This perspective paper aims to delve into the promises and challenges of advancing human learning in the age of GenAI. By integrating human-centred theories of learning and instruction, we emphasise the importance of designing AI-driven educational tools that prioritise the needs of learners in contemporary societies. We elaborate on how this technology can transform learning and teaching practices while remaining critical of the ethical and practical challenges it poses, contributing to a future research agenda for investigating human-AI interaction and the adoption of GenAI as a tool for learning (Fig. \ref{fig: overview}). 

\section{Promises}\label{sec-pro}

\begin{figure}[ht]
    \centering
      \includegraphics[width=.99\textwidth]{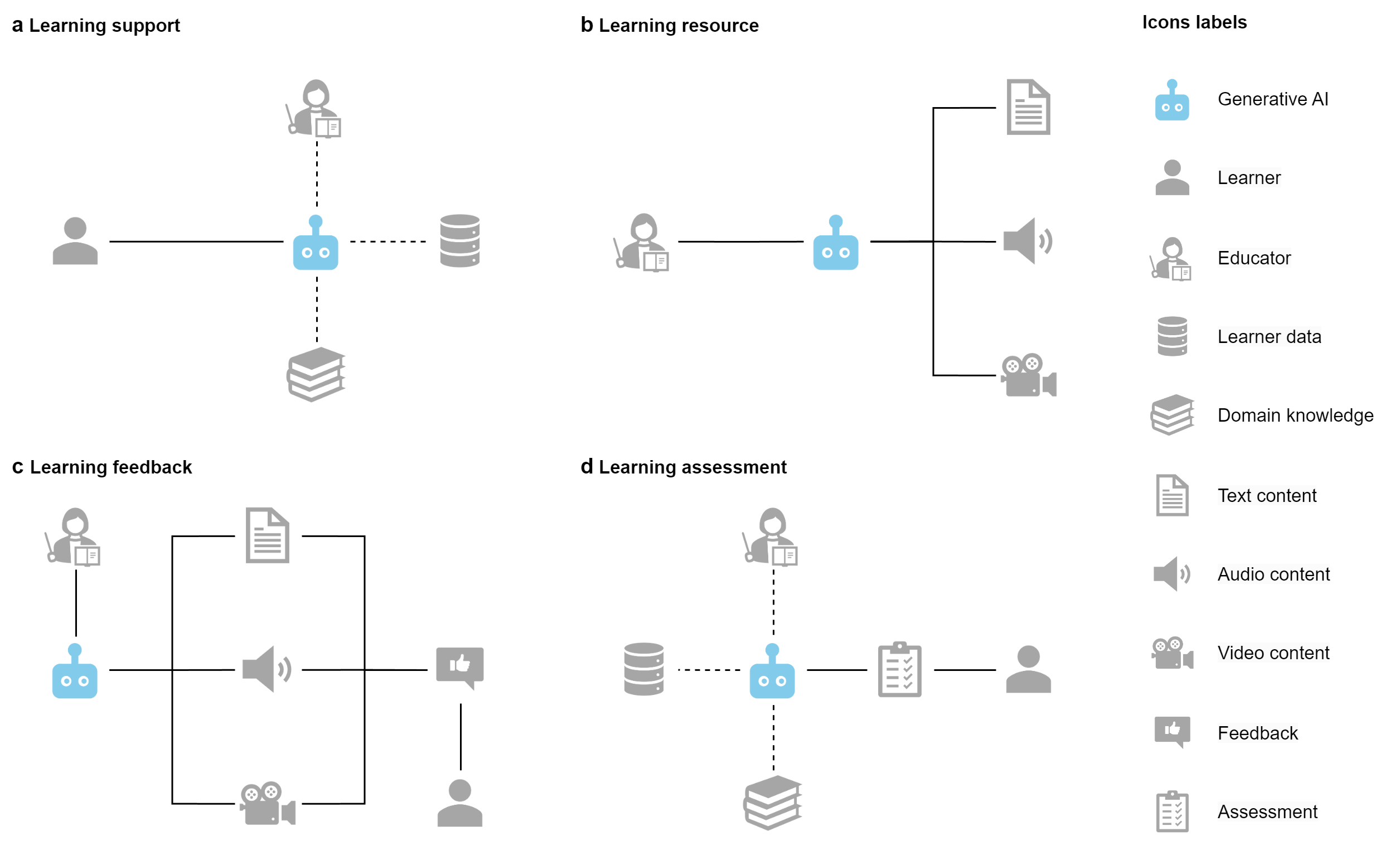}
    \caption{\textbf{Examples of human-AI interactions in human learning.} \textbf{a,} Learners receive personalised and adaptive support from generative AI tutors, which are co-designed with educators and have access to prior learner data and domain knowledge. \textbf{b,} Educators use generative AI to create multimodal learning resources, incorporating text, audio, and video content. \textbf{c,} Educators collaborate with generative AI to deliver multimodal feedback to learners. \textbf{d,} Generative AI agents use input requirements from educators, prior learner data, and domain knowledge to create assessment activities that evaluate learners.
}
    \label{fig: examples}
\end{figure}

GenAI promises to transform human learning by scaling personalised support, diversifying learning resources, enabling timely feedback, and innovating assessment methods. The realisation of these promises depends on the roles and interactions GenAI has with learners and educators (Fig. \ref{fig: examples}). Specifically, GenAI technologies can act as cognitive facilitators within learners' Zone of Proximal Development, providing adaptive support at scale. GenAI can also enrich learning experiences by assisting in the creation of diverse multimedia resources. In feedback, GenAI systems offer timely and multimodal insights, surpassing traditional methods in depth and efficiency. For assessment, GenAI enables adaptive and authentic evaluations using generative agents and multimodal models. The following sections explore each of these promises, illustrating their potential to transform the delivery, cultivation, and evaluation of human learning.

\subsection{Learning Support}\label{sec-pro-learning}

The unique contribution of GenAI, particularly LLMs, to learning support lies in its scalability and adaptability. GenAI can function as a master teacher at scale, providing personalised and adaptive support to a wide range of learners across various subjects and languages. Unlike conventional intelligent tutoring systems that require extensive knowledge engineering to design rule-based responses \cite{mousavinasab2021intelligent}, GenAI can achieve superior and more naturalistic interactions, such as personalised feedback, adaptive questioning, and conversational engagement, with minimal prior training. These enhanced interactions facilitate more effective and intuitive tutoring experiences, making the learning process more engaging and tailored to individual student needs \cite{kasneci2023chatgpt}. This capability holds the potential to democratise access to high-quality learning support, making it accessible to learners globally.

GenAI's role aligns with Vygotsky's sociocultural theory of learning, where more knowledgeable others guide learners within their Zone of Proximal Development \cite{vygotsky1978mind}. By integrating novel technologies like ChatGPT into intelligent tutoring systems, GenAI can offer personalised and adaptive support based on each learner's unique knowledge \cite{joksimovic2023opportunities}. These language models have demonstrated remarkable proficiency in processing semantic and contextual information \cite{chang2024survey}, a key aspect of their effectiveness as a tool for learning. By accurately interpreting and responding to the linguistic and contextual nuances in learners' queries, LLMs ensure that the learning experience is interactive and thought-provoking. Rather than merely dispensing solutions, they can be used to encourage learners to engage cognitively with the material. This engagement is achieved by prompting students to think critically, unpack problems, and understand underlying concepts.

A representative case of how GenAI can support learning comes from Khan Academy's "Khanmigo" chatbot, powered by GPT-4 and designed to assist with mathematical queries \cite{khanacademy2023}. Khanmigo exemplifies the shift from providing direct answers to a more nuanced, guided learning approach that offers constructive feedback and step-by-step instruction. For example, when students present Khanmigo with a problem on fractions, it guides them through the underlying concepts of denominators and numerators, encouraging them to apply these concepts to solve the problem through a series of guiding questions. Khanmigo functions as a facilitator, aligning with the principles of inquiry-based learning \cite{lee2012inquiry}, a human-centred learning theory that emphasises the importance of active learning through inquiry. This theory encourages students to ask questions, explore, and engage deeply with the learning material to develop deep knowledge. This iterative methodology reflects Vygotsky's emphasis on the importance of the learning journey over the destination by fostering deep conceptual comprehension and retention \cite{lee2012inquiry,vygotsky1978mind}. By engaging learners in a dialogic process, GenAI-driven systems such as Khanmigo aim to enhance learners' critical thinking and problem-solving skills.

Despite the promising design of systems like Khanmigo and students' positive attitudes towards using such technologies for personalised learning support \cite{chan2023students}, it is important to acknowledge the current limitations in empirical evidence regarding their short- and long-term impacts on learning outcomes \cite{hennessy2024bjet}. Emerging evidence indicates that the impact on learning engagement, agency, and performance can paint a complicated and mixed picture (e.g., lack of learning gains after removing GenAI supports) \cite{darvishi2024impact,nie2024gpt}. Therefore, further research is needed to substantiate GenAI's long-term benefits to human learning. This includes conducting longitudinal and randomised controlled studies that compare the effectiveness of GenAI tutoring with conventional rule-based tutoring systems over several academic terms and across different subjects to contextualise its impacts within various educational settings.

\subsection{Learning Resource}\label{sec-pro-resource}

Effective learning relies on the quality and diversity of resources, yet developing high-quality materials is often time-consuming and resource-intensive. GenAI promises to ease this burden by creating diverse and engaging content, fostering curriculum innovation and enhancing learning experiences. Studies on human-AI collaboration indicate that co-creating content with GenAI can meet diverse learning needs, providing students with relevant and accessible materials to support their individual paths efficiently and creatively \cite{molenaar2022towards,ji2023systematic,yang2021surveying}. For instance, early explorations have shown that GPT-4 can automatically generate instructional materials, such as explanations of programming concepts, examples, and quiz questions, thereby boosting learner engagement and satisfaction \cite{pesovski2024generative}. Additionally, GPT-4 has demonstrated proficiency in generating college-level biology questions for lower levels of Bloom’s Taxonomy (e.g., remember and understand) but struggles with higher levels (e.g., apply and create) \cite{hwang2024towards}. These findings suggest that while GenAI can produce learning resources, educators' expertise remains crucial for ensuring the accuracy, relevance, and pedagogical soundness of the material. This highlights the need for a human-AI collaborative approach to create meaningful resources that meet diverse learning objectives and learner needs.

GenAI can also enrich learning resources by generating interactive activities, multimedia content, and real-world problem-solving scenarios. Text-to-image models like Stable Diffusion, Midjourney, and DALL-E \cite{radford2021learning, chiu2023impact} enable educators to create visual learning materials from textual content. These tools can foster students' creative thinking by engaging them in activities such as using AI to generate images. For instance, students can create imaginative visuals with AI and write inspired diaries based on these images, a practice found to reduce gender disparities in interest in art during Science, Technology, Engineering, the Arts and Mathematics (STEAM) classes \cite{lee2023prompt}. This innovative approach has also been shown to enhance primary school students’ extrinsic motivation, problem-solving awareness, critical thinking, and learning performance in ancient Chinese poetry \cite{chen2024progressive}. Similarly, text-to-video generation tools like Runway's Gen-3 Aphla and OpenAI's Sora can support educators in creating video narratives from textual content, further diversifying learning modalities. This capability is particularly valuable for teaching students with specific disabilities, such as providing multisensory instruction to students with dyslexia \cite{long2007supporting}. A preliminary study found no significant differences in learning gains and perceived experiences between GenAI-generated videos with synthetic instructors and traditional recorded instructor videos, suggesting that GenAI could make high-quality learning resources more accessible globally \cite{leiker2023generative}.

By offering a range of pedagogical possibilities through efficient and diversified resource development, GenAI can help educators create more dynamic and engaging learning environments. This enables learners to interact with content in more informed, creative, and personalised ways. Such an approach aligns with constructivist learning principles, which emphasise the importance of learners actively constructing knowledge through exploration and interaction \cite{bada2015constructivism}. However, more research is needed to balance this integration of GenAI in developing resources for human learning \cite{molenaar2022towards,tavakoli2022hybrid}, such as determining the optimal level of automation versus human control, the extent of expert validation required, and the degree of alignment with learning objectives.

\subsection{Learning Feedback}\label{sec-pro-feedback}

Another promise of GenAI in supporting human learning is its potential to provide timely, specific, and constructive feedback, a key element of high-quality instruction and essential for effective learning \cite{pardo2019using, lim2021changes, hattie2007power}. Providing detailed feedback regularly is laborious and time-consuming, adding to educators' workloads, especially since students perceive timely feedback as the most effective \cite{poulos2008effectiveness}. GenAI can assist by analysing student work and delivering instant, personalised feedback with minimal prior training. For example, a recent study found that ChatGPT generates more in-depth and fluent feedback, coherently summarising students' performances compared to human educators \cite{dai2023can}. This AI-generated feedback also includes process-focused elements, which are more effective in shaping learning strategies \cite{hattie2007power}. 

Emerging studies show similar benefits in various learning contexts, such as formative feedback in secondary school essay writing \cite{steiss2024comparing,meyer2024using}, programming assignments in introductory computer science courses \cite{zhang2024students,liang2024towards}, and collaborative second language writing \cite{wiboolyasarin2024synergizing}. GenAI-generated feedback has led to enhanced task performance and positive experiences \cite{meyer2024using,zhang2024students,wiboolyasarin2024synergizing}. Additionally, chatbots powered by GenAI models with natural and visual language understanding capabilities (e.g., GPT-4 with Vision and Gemini 1.0 Pro) can help students navigate and comprehend insights from learning analytics dashboards \cite{yan2024vizchat}, which combine data, analytics, and visualisations to provide feedback on learning processes and outcomes \cite{matcha2019systematic}. These chatbots could facilitate dialogic feedback, which is associated with improved learner productivity and engagement \cite{yang2013feedback}.

GenAI could also expand feedback delivery beyond text and graphics to include narrated audio and video, addressing the scalability challenges of these formats and leveraging their benefits for enhanced feedback efficiency and student engagement \cite{dawson2023technology}. For example, by combining 3D diffusion models \cite{wang2023rodin} and text-to-speech models \cite{le2024voicebox}, educators can create digital avatars to convey feedback through a narrated voice rather than text alone. This diversity in feedback modalities can increase engagement and effectiveness \cite{dawson2023technology}. Prior research indicates that audio and video feedback is often perceived as more personal and dynamic, enhancing understanding and engagement compared with traditional written feedback \cite{mccarthy2015evaluating, orlando2016comparison, henderson2015video}. The integration of GenAI technologies promises to facilitate timely and multimodal feedback, providing more informative feedback and fostering improved effectiveness and engagement in the learning process. However, it is essential to evaluate the cost and benefits, as these models, especially video generation models, require high computational power, potentially widening the inequality in learning opportunities.

\subsection{Learning Assessment}\label{sec-pro-assessment}

GenAI is transforming the assessment of learning, shifting from traditional, often onerous methods to more adaptive and authentic processes \cite{swiecki2022assessment}. Central to this shift is GenAI's potential to create personalised and adaptive assessments, enhancing the understanding of each student's needs and progression. This is enabled by advancements in generative agents -- autonomous and adaptive AI entities that operate independently, pursuing objectives without continuous user interaction, as exemplified by tools like AutoGen (preprint) \cite{wu2023autogen}. These agents exhibit human-like cognitive and metacognitive abilities, including task planning, situational assessment, progress monitoring, and collaborative efforts among agents. For instance, a group of 25 generative agents in a dynamic sandbox environment successfully organised and conducted a Valentine's Day celebration based on a single user input \cite{park2023generative}. By leveraging a similar agent architecture, encompassing observation, planning, and reflection, and integrating these with process-centred methodologies (e.g., modelling self-regulated learning from learners' digital traces \cite{fan2022towards}) from the field of learning analytics \cite{allen_natural_2022,gavsevic2022towards}, learning scientists and researchers can develop generative agents capable of autonomously evaluating human learning. These agents can identify areas of knowledge deficiency and provide tailored learning resources and adaptive assessments.

Recent educational technology studies highlight the potential of automated assessments through multi-agent frameworks that leverage multiple LLM agents. These GenAI systems are being used to grade coding assignments in online learning \cite{lagakis2024evaai}, conduct cognitive assessments to identify students' strengths and weaknesses according to Bloom's taxonomy in e-learning environments \cite{shahzad2024multi}, and assess educators' mathematical content knowledge for professional development programs \cite{yang2024content}. These applications demonstrate strong potential for generalisability, precision, and dependability.

GenAI also holds promise for advancing authentic assessments \cite{swiecki2022assessment}. It can enhance learning tasks in both virtual and physical simulations to more accurately mirror real-world situations, making assessments more meaningful and contextualised. Previous studies have shown the effectiveness of combining LLMs with knowledge graphs to create virtual standard patients, aiding the training and evaluation of medical students' diagnostic skills \cite{song2022intelligent}. Knowledge graphs are structured representations that integrate diverse data sources, providing a comprehensive understanding of a domain \cite{ji2021survey}. When used with LLMs, they can simulate complex learning and assessment scenarios requiring critical thinking and problem-solving skills, such as in driving education \cite{rehm2024virtual}, programming education \cite{jin2024teach}, and laboratory safety courses \cite{yang2023developing}. Integrating multimodal generative models, such as GPT-4 Vision for text and image generation, Meta's Voicebox for audio creation from text, and generative adversarial networks for digital avatar production, can further enhance the authenticity of simulated assessment environments \cite{thanh2023race}. These enhancements allow students to interact naturally and perform procedural actions as if they were in real professional settings, a concept proven effective in virtual internships \cite{chesler2015novel} and healthcare simulations \cite{cant2010simulation}. However, much effort is required to develop valid and reliable behavioural and psychological indicators in these novel assessment settings to accurately capture genuine human learning.

\section{Challenges}\label{sec-challenges}

Amidst GenAI's promises, formidable challenges confront learners and educators alike and raise critical moral and ethical concerns about integrating such technology into human learning. These challenges involve GenAI technologies' imperfections, the ethical dilemmas of transparency, privacy, equality, and beneficence, and the disruption of assessment practices. The following sections elaborate on each of these challenges. 

\subsection{GenAI's Imperfections}\label{sec-challenges-imperfect}

As GenAI technologies become increasingly integrated into learning support, resource generation, feedback, and assessment, it is imperative to address the risks posed by hallucinations \cite{maynez2020faithfulness}. Hallucinations occur when there are mismatches in training data or complexities in language generation tasks, resulting in outputs that may not align with factual information \cite{ji2023survey}. The probabilistic nature of LLMs and diffusion models further limits their utility due to inherent instabilities and potential biases in their training data \cite{carlini2021extracting}. For instance, ChatGPT often fails tasks easily solved by humans, such as reasoning tasks requiring real-world knowledge, logic, mathematical calculations, and distinguishing between factual and fictive information. Consequently, it sometimes provides fabricated facts \cite{borji2023categorical} (preprint). These inaccuracies can undermine GenAI's reliability as a learning tool, potentially outweighing its promises (Section \ref{sec-pro}).

Emerging studies indicate that hallucinations in GenAI can occur with non-negligible frequency, increasing with the complexity and specificity of queries posed to the AI \cite{ji2023survey}. GenAI may perform reasonably well with generic questions (e.g., What are Newton's laws of motion?) but is more prone to errors with nuanced, context-specific, time-sensitive, or highly technical information \cite{chelli2024hallucination}. The lack of transparency in GenAI's decision-making process complicates identifying when and why these hallucinations occur \cite{sahoo2024addressing,ji2023survey}. Relying solely on GenAI for learning content creation and curriculum development without validation could introduce inaccuracies, misleading both educators and students. Similarly, GenAI-generated feedback or assessments based on incorrect information could misguide students' learning processes, leading to misconceptions or a lack of understanding of key concepts. 

Addressing these challenges requires an interdisciplinary effort. Educators should adopt a balanced and proactive approach, teaching learners to critically evaluate AI-generated content by cross-referencing with reliable sources, questioning plausibility, and recognising signs of hallucination. These steps are essential for cultivating AI literacy \cite{ng2021conceptualizing}, as discussed further in Section \ref{sec-future-literacy}. Additionally, designing and optimising the interface of educational technologies to highlight potential hallucinations requires collaboration among learning scientists, human-computer interaction researchers, and technology providers \cite{leiser2023chatgpt,sahoo2024addressing}. Such a collaborative approach is essential to empower learners to deal with the imperfections of GenAI both intrinsically, by developing critical thinking skills, and extrinsically, by leveraging improved technological interfaces that signal potential inaccuracies.

\subsection{Ethical Dilemmas}\label{sec-challenges-ethics}

Adopting GenAI to support human learning raises several ethical issues, notably in areas such as transparency, privacy, equality, and beneficence. A key concern is the transparency of AI-generated solutions, as highlighted in a recent systematic literature review \cite{yan2023practical}. The review found that a majority (92\%) of GenAI tools currently used for supporting learning practices, particularly those based on LLMs, are transparent only to AI experts, not to educators, students, or other stakeholders. This lack of transparency is problematic as it obscures the understanding of AI functionalities and potential flaws from those directly impacted by these technologies \cite{schneider2023towards}. The primary cause of this transparency gap is the absence of human-in-the-loop elements in prior research, such as involving educators and students in the development and evaluation of GenAI-powered educational technologies. This aligns with the growing emphasis on developing explainable and human-centred AI, underscoring the essential role of stakeholder involvement in crafting impactful and meaningful educational technologies \cite{khosravi2022explainable, yang2021human}.

To achieve personalisation in learning support, resource generation, feedback, and assessment using GenAI, learners' personal data must be provided to these models. However, privacy concerns can reduce learner participation \cite{short2014critical, mutimukwe2022students}. These concerns are prominent due to the lack of clear consent strategies and data protection measures surrounding GenAI in supporting human learning \cite{yan2023practical}. Using learner-generated data without explicit consent or adequate anonymisation raises serious issues about exposing sensitive information \cite{brown2022does}. For instance, researchers conducted a divergence attack on ChatGPT, compromising its security and causing it to output original training data containing personally identifiable information \cite{nasr2023scalable} (preprint). Although OpenAI has addressed this vulnerability, potential data breaches from unforeseen attacks remain a concern \cite{winograd2023loose, yao2024survey}. This issue is particularly troubling given the resources required for GenAI to unlearn information once private data has been used for model training, especially for large, commercial, and proprietary models \cite{winograd2023loose}.

Regarding equality, there is an evident disparity in language representation and accessibility of GenAI models. While advancements have been made in non-English languages for LLMs and speech diffusion models \cite{le2024voicebox, chang2024survey}, the predominance of English-based AI solutions perpetuates a bias towards Western, Educated, Industrialized, Rich, and Democratic (WEIRD) societies \cite{yan2023practical}. This imbalance raises concerns about the global applicability and fairness of these technologies, potentially intensifying existing inequalities and the digital divide in learning opportunities \cite{pugh2021say}. 

Finally, beneficence is a critical ethical principle that must be addressed. Several studies highlight the risks of underperforming or biased AI models, which can negatively impact human learning and perpetuate systemic biases, such as gender, racial, and social class biases \cite{sha2021assessing, merine2022risks}. Strategies like balanced sampling and cautionary labelling have been proposed \cite{sha2022leveraging, sha2022bigger}, but the opaque nature of many generative models makes ensuring fairness and accuracy challenging, potentially violating the principle of beneficence \cite{wu2022analysis}. While model alignment is often implemented to prevent GenAI from producing toxic content, recent evidence suggests that adversarial attacks using specific prompts can undermine these measures \cite{yao2024survey}. Such attacks could facilitate cheating, promote biased views, or expose students to offensive language \cite{tlili2023if}. These issues could disrupt learning, compromise safety and inclusivity, and cause psychological harm, eroding trust in educational technologies. These ethical challenges underscore the need for rigorous and multifaceted ethical considerations in deploying GenAI and the urgency of establishing regulations, such as the EU AI Act \cite{eu2023aiact}.

\subsection{Disruption of Assessment}\label{sec-challenges-assessment}

GenAI poses significant challenges to conventional learning assessment methodologies. Traditionally, assessments have focused on evaluating learning products, such as essays, to measure outcomes \cite{swiecki2022assessment}. However, GenAI's ability to produce high-quality, human-like responses calls into question the validity of these approaches \cite{mao2023generative}. A central issue is distinguishing between a learner's work and AI-generated output. GenAI, particularly LLMs like ChatGPT and Llama 3, can generate responses that closely mimic human reasoning and writing styles, making it difficult to discern the origin of the work \cite{li2023can}. 

A performance paradox arises when tasks are completed with GenAI assistance. A recent randomised controlled experiment found that while GenAI tools can help students achieve better performance, removing this support significantly lowers their performance \cite{darvishi2024impact}. This suggests that GenAI may create an illusion of improved learning without developing essential skills, such as self-regulated learning. Thus, we must ask: Who and what are we actually assessing? This dilemma extends beyond detecting AI-generated content to reconsidering the purpose of assessment in learning.

The challenge is further compounded when considering the learning process itself. GenAI's ability to interact with computational systems means even the learning process can be imitated or augmented by AI. Preliminary work on multimodal GenAI agents \cite{yang2023appagent} (preprint) has shown these agents can operate smartphone applications, generating digital trace data while executing user requests. This AI-generated data could impede existing learning analytic methods that rely on such data to model the learning process \cite{viberg2018current}. This issue blurs the line between human cognition and AI-augmented cognition \cite{siemens2022human}, complicating the assessment of skills traditionally seen as exclusively human, such as critical thinking, problem-solving, and creativity \cite{mao2023generative}. 

Consequently, we must reconsider the purpose of learning assessment across different educational stages. Assessing human cognition and metacognition remains essential for K-12 education, as young learners continue developing fundamental skills. In higher education, prioritising the evaluation of human-AI hybrid cognition and metacognition could be crucial for preparing learners for an AI-integrated workforce \cite{jarvela2023hybrid}. This shift demands rethinking assessment strategies to accommodate the collaborative nature of learning in the presence of AI.

\section{Needs}\label{sec-future}

Within GenAI's promises and challenges, three pivotal needs must be addressed for effective integration into human learning: cultivating AI literacy among learners and educators, prioritising evidence-based decision-making, and ensuring methodological rigour in research using GenAI. These needs aim to foster a balanced integration that enhances human abilities and ensures a synergistic relationship between GenAI and human development.

\subsection{AI Literacy}\label{sec-future-literacy}

Cultivating AI literacy is essential to ensuring the effective, responsible, and ethical use of GenAI technologies to support human learning \cite{long2020ai,ng2021conceptualizing}. This need extends beyond learners to include educators, policymakers, and administrators, who are integral to the design, delivery, and facilitation of learning experiences. AI literacy encompasses a basic understanding of how AI systems function but also an intimate awareness of their potential impact, ethical considerations, and limitations \cite{ng2021conceptualizing}. The absence of AI literacy can lead to severe consequences. For instance, the New York Times reported that a lawyer using ChatGPT for a court filing was unaware of fabricated citations generated by the AI, resulting in a breach of professional ethics and legal standards \cite{nytimes2023avianca}. One must ask: What if educators unknowingly provided students with AI-generated learning resources that contained fabricated content? Such actions could erode trust and integrity in education systems, misleading students and compromising their learning quality. 

These concerns highlight the critical need to cultivate AI literacy. A recent study indicates that human users often prefer AI-generated content for its comprehensiveness and well-articulated language style, despite its inaccuracies \cite{kabir2024stack}. As GenAI's propensity to hallucinate remains challenging to address at the foundational model level \cite{ji2023survey}, understanding its limitations and identifying potential pitfalls will be crucial for preparing individuals to live, learn, and work with GenAI in the 21st century. This requires adopting AI literacy models, practices for their development, and measurement approaches. Institutions, policymakers, and researchers must focus on AI literacy as a key learning objective to ensure that educators, students, administrators, and even parents are not merely consumers of AI technology but also informed participants in its evolution and application.

\subsection{Evidence-Based Decision Making}\label{sec-future-impact}

The integration of GenAI into human learning promises to enhance experiences and outcomes (as highlighted in Section \ref{sec-pro}). However, adopting these technologies requires a commitment to evidence-based decision-making. This necessitates a collaborative effort among researchers, practitioners, and policymakers to generate robust evidence guiding the effective and responsible use of AI in learning practices. By working together, these stakeholders can ensure GenAI deployment aligns with learning goals and supports the development of essential cognitive and metacognitive skills.

Encouraging the use of GenAI to support human learning requires a nuanced understanding of its benefits and limitations. For instance, while GenAI can improve the efficiency of information processing and retrieval, there is a risk of fluency bias, where learners may overestimate their understanding due to the ease of cognitive information processing \cite{bjork2013self,kabir2023answers} (preprint). Similarly, reliance on GenAI for creative and problem-solving tasks could weaken these critical skills, fostering a dependency that may hinder innovation and original thought \cite{rafner2023creativity, shneiderman2020human, UNESCO_2023}.

To mitigate these risks and maximise GenAI's benefits, it is imperative to foster partnerships among researchers, practitioners, and policymakers. These collaborations can produce evidence that informs learning and teaching practices, ensuring that GenAI enhances rather than replaces human cognitive, metacognitive, and creative processes. By prioritising evidence-based decision-making and stakeholder collaboration, we can leverage GenAI's advantages in educational environments while promoting deep learning, creativity, and problem-solving abilities among learners.

\subsection{Methodological Rigour}\label{sec-future-rigour}

Building on discussions of evidence-based decision-making, it is crucial to emphasise methodological rigour in applying GenAI technologies within human learning research. As these technologies evolve, human learning researchers and scientists must adapt and refine their methodologies to accurately assess the impact of these tools on teaching and learning processes. GenAI's capabilities, such as passing the United States Medical Licensing Exam \cite{kung2023performance}, completing exams at the University of Minnesota Law School \cite{choi2021chatgpt}, and solving queries from Wharton School of Business tests \cite{terwiesch2023would}, underscore its potential. However, the excitement must be tempered with caution to avoid overestimating effectiveness due to methodological shortcomings. A notable example is a preprint study claiming GPT-4, with prompt engineering, could achieve perfect scores in the MIT Mathematics, Electrical Engineering, and Computer Science curriculum \cite{zhang2023exploring}. This study \cite{zhang2023exploring}, initially attracting widespread attention, was later retracted due to methodological concerns, including data set contamination, over-reliance on GPT-4 for accuracy assessment, and ambiguities in manual verification of results \cite{chowdhuri2023uom} (preprint). This incident underscores the need for rigorous methodological standards, likely requiring new approaches in evaluating GenAI technologies.

To address these challenges, it is essential to establish standards for appraising the quality of evidence on GenAI's impact on learning processes, outcomes, and experiences \cite{hennessy2024bjet}. In the medical field, tools such as the Cochrane Risk of Bias Tool and ROBINS-I are used to assess study quality. Given the distinct methodological requirements introduced by GenAI, including various prompting engineering strategies and retrieval generation techniques, it is crucial to establish specific quality standards and evaluation tools. These requirements go beyond conventional methodologies used in human learning research. For example, using GenAI to generate physics practice questions might involve retrieval methods that limit the AI to sourcing content solely on Newton's laws of motion and crafting prompts specifying complexity level, target student grade, and desired question format (e.g., multiple-choice, short answer, or problem-solving). By working collaboratively, the human learning research community can create a robust framework for evaluating evidence, ensuring a solid foundation for future policies and practices. This effort will enable researchers, practitioners, and policymakers to build on reliable, valid, and generalisable findings, fostering the responsible and effective integration of GenAI technologies into learning.

\section{Conclusion and Future Directions}\label{sec2}

As we look toward the next decade, powerful AI tools are set to become integral to our society, transforming how we learn, work, and live \cite{lorenz2023initial}. GenAI technologies could permeate every aspect of human learning. Imagine students collaborating with AI agents designed to mimic certain personality traits to help students learn about leadership and teamwork, engaging in debates with digital twins of Socrates, Plato, and Aristotle to explore ancient Greek philosophy, learning impressionist painting techniques from a humanoid robotic mentor modelled after Claude Monet, and visualising Einstein's special theory of relativity in virtual realities. All this could occur while receiving personalised support from a GenAI tutor hosted on a wearable device. This integration necessitates a dual approach to learning: educating ourselves both about and with GenAI, while continuing to develop critical thinking, problem-solving, self-regulation, and reflective thinking skills. These skills are crucial for maintaining cognitive and metacognitive autonomy as AI becomes embedded in our daily lives.

Understanding the relationship between GenAI and human cognition, metacognition, and creativity is essential for maximising its potential as a learning tool. This understanding will enhance the effectiveness of AI-driven educational tools and ensure human ingenuity is preserved amidst technological advancement. Key research questions include: How can we promote human-AI interaction to maximise learner agency? What behavioural indicators can reliably capture cognitive and metacognitive processes during AI-assisted learning? How can we assess learning to reflect genuine knowledge and skill development rather than an AI-created performance illusion? What strategies can prevent over-reliance on AI, ensuring humans remain primary agents of critical thinking and problem-solving?

Educators are pivotal in integrating AI tools to enhance traditional teaching methods. We anticipate a shift in educators' roles, with GenAI reducing the burden of knowledge dissemination, allowing teachers to focus on deeper connections with students as mentors and facilitators. This transition requires educators to adopt new pedagogical paradigms that leverage AI to foster intellectual and emotional growth. They must become proficient in AI literacy, effectively integrate AI tools into their teaching, and remain vigilant about potential pitfalls, such as GenAI's imperfections and the risk of student over-reliance on AI. Balancing AI use with activities promoting human creativity, critical thinking, and social interaction is crucial to ensure AI augments rather than replaces human educators. Educational institutions must invest in ongoing professional development and support systems to help teachers manage techno-stress and workload burdens from adopting new technologies.

Policymakers and technology companies should consider: How can we ensure accountability for AI tools used in human learning, and who should be responsible for their outcomes? What ethical guidelines should govern AI tools in educational settings? How can we design and implement AI learning tools to promote equality and inclusivity?

We argue that human-centred theories of learning and instruction must be integrated with GenAI to ensure these technologies enhance rather than detract from human learning. This involves developing AI systems that support and elevate human cognitive capacities. By fostering a learning environment that harmonises technology with theoretical approaches, we can promote personal growth and the development of adaptive skills and knowledge needed to navigate the rapid changes in the age of AI. A united effort among researchers, policymakers, technology companies, and educators is essential to fully leverage GenAI's potential in advancing human learning. By addressing these critical questions and considerations, we can ensure that GenAI becomes a powerful ally in the pursuit of knowledge and innovation, rather than a crutch that undermines our intellectual abilities.

\begin{table*}[htbp]
\centering
\caption{Defining human learning concepts}
\label{tab:learning}
\begin{tabular}{|p{13cm}|}
\hline
\textbf{Human learning.} Human learning is the process through which individuals acquire new knowledge, skills, attitudes, or values. \\
\hline
\textbf{Education.} Education is the structured process of teaching and learning, typically occurring in institutional settings such as schools, universities, and training programs. It aims to develop learners' intellectual, social, and emotional capacities, preparing them for personal and professional success. \\
\hline
\textbf{Constructivist learning.} Constructivist learning is a theory that posits learners construct their own understanding and knowledge of the world through experiences and reflecting on those experiences. It emphasises active engagement, exploration, and the application of knowledge in meaningful contexts. \\
\hline
\textbf{Inquiry-based learning.} Inquiry-based learning refers to pedagogical approaches in which learners use scientific methods to build knowledge through formulating hypotheses, conducting experiments, and making observations to discover causal relations. It emphasises problem-solving, active participation, and self-directed learning, facilitating inductive and deductive reasoning. The knowledge gained is usually new to the learner, fostering deep understanding through exploration. \\
\hline
\textbf{The zone of proximal development.} The zone of proximal development represents the gap between what a learner can do on their own and what they can achieve with help from a skilled guide. It encompasses tasks that are currently beyond the learner’s independent capabilities but can be completed with assistance and guidance. \\
\hline
\textbf{Bloom’s taxonomy.} Bloom's Taxonomy is a framework used to categorise learning goals and objectives and provides a structured approach to designing learning content and assessing learning outcomes. It classifies cognitive skills into a hierarchy from basic to advanced: remember, understand, apply, analyse, evaluate, and create. \\
\hline
\textbf{Learning analytics.} Learning analytics involves the collection, measurement, and analysis of data about learners and their contexts to understand and optimise learning and the environments in which it occurs. It uses data-driven insights to inform educational decision-making and enhance learning outcomes. \\
\hline
\textbf{Feedback.} Feedback refers to responses related to a learner’s performance or understanding, information to them about the correctness of task solutions or providing them with content-related or strategic assistance and information about their processing. The German psychologist Lipowsky identifies feedback as one of the nine hallmarks that characterise high-quality instruction. \\
\hline
\textbf{Authentic assessment.} Authentic assessment refers to evaluation methods that require learners to apply their skills and knowledge to real-world tasks and problems. It aims to assess learners' abilities in contexts that are relevant and meaningful, providing a more accurate measure of their competencies. \\
\hline
\textbf{Intelligent tutoring system.} An intelligent tutoring system is a computer-based system or tool created to mimic human tutoring. It provides learners with immediate, personalised instruction or feedback, often functioning autonomously without requiring direct intervention from a human teacher. \\
\hline
\textbf{Digital twins.} Digital twins are virtual replicas of physical entities, such as objects, systems, or processes. In education, digital twins can simulate real-world scenarios, providing learners with immersive and interactive experiences to enhance understanding and skill development. \\
\hline
\end{tabular}
\end{table*}

\begin{table*}[htbp]
\centering
\caption{Glossary of artificial intelligence terms}
\label{tab:ai}
\begin{tabular}{|p{13cm}|}
\hline
\textbf{Generative artificial intelligence.} Generative artificial intelligence refers to AI systems designed to create new content, such as text, images, or music, by learning patterns from existing data. These systems can produce outputs that are novel and relevant, often mimicking human creativity. \\
\hline
\textbf{Large language model (LLM).} An LLM is a computational model known for its ability to perform general-purpose language generation and various natural language processing tasks like classification. LLMs acquire these abilities by learning statistical relationships from vast text datasets during an intensive self-supervised and semi-supervised training process. They can generate text by taking inputs and repeatedly predicting the next word, a form of generative AI. \\
\hline
\textbf{Diffusion model.} A diffusion model is a type of probabilistic model that generates data by simulating the gradual transformation of noise into coherent data points. These models use a series of iterative steps to refine random noise into structured outputs, such as images or text. Diffusion models have shown promise in generating high-quality, realistic data and are used in applications like image synthesis, text generation, and other creative tasks. \\
\hline
\textbf{Knowledge graph.} A knowledge graph is a structured representation of information that captures relationships between various entities, such as objects, events, or concepts. It organises data into nodes (representing entities) and edges (representing relationships between entities), enabling complex queries and inferences. Knowledge graphs are used in applications like search engines, recommendation systems, and natural language understanding to provide context-aware and semantically rich insights. They help in connecting and integrating diverse data sources, thus enhancing the ability of AI systems to understand and reason about the world. \\
\hline
\textbf{AI Agent.} An AI agent is an autonomous and adaptive AI entity that operates independently, pursuing objectives without continuous user interaction. \\
\hline
\textbf{AI literacy.} AI literacy involves understanding the fundamental concepts and capabilities of artificial intelligence, as well as its implications and ethical considerations. It encompasses the skills needed to interact with AI systems effectively and to critically evaluate their outputs. \\
\hline
\textbf{Hallucinations.} In the context of AI, hallucinations refer to instances where generative models produce outputs that are incorrect, nonsensical, or fabricated. These errors can occur due to the model's limitations or biases in the training data. \\
\hline
\textbf{Divergence attack.} A divergence attack is a method used to exploit weaknesses in AI models, causing them to deviate from their intended behaviour. This can result in the model generating harmful or unintended outputs, potentially exposing sensitive information or producing biased content. \\
\hline
\textbf{Model alignment.} Model alignment involves ensuring that AI systems' behaviours and outputs are consistent with human values and intended goals. It includes efforts to make AI systems safe, reliable, and ethical in their operations. \\
\hline
\end{tabular}
\end{table*}

\bibliography{sn-bibliography}

\section*{Acknowledgments}

This study was in part supported by grants from the Australian Research Council (grant agreement numbers DP220101209 and DP240100069 to D.G.). L.Y.'s work is fully funded by the Digital Health CRC (Cooperative Research Centre). D.G.'s work was, in part, supported by the DHCRC and Defense Advanced Research Projects Agency (DARPA) through the Knowledge Management at Speed and Scale (KMASS) program (HR0011-22-2-0047). The DHCRC is established and supported under the Australian Government's Cooperative Research Centres Program. The U.S. Government is authorised to reproduce and distribute reprints for Governmental purposes notwithstanding any copyright notation thereon. The views and conclusions contained herein are those of the authors and should not be interpreted as necessarily representing the official policies or endorsements, either expressed or implied, of DARPA or the U.S. Government.

\section*{Competing interests}
The authors declare no competing interests.

\end{document}